\documentstyle[a4,11pt,epsf]{article}

\epsfverbosetrue
\newcommand{\beq}{\begin{equation}}
\newcommand{\eeq}{\end{equation}}   

 0

\begin{document}

\begin{titlepage}
 
\vspace{5mm}
 
\begin{center}
{
 \huge 
    The Structure of the Aoki Phase
  \\[3mm]
       at Weak Coupling
}
\\[15mm]
{\bf 
R. Kenna,
C. Pinto
and 
J.C. Sexton,
\\
School of Mathematics, Trinity College Dublin, Ireland
} 
\\[3mm]~\\ 
January 2001
\end{center}
\begin{abstract}
A  new method to determine the phase diagram of certain 
lattice fermionic field theories in the weakly coupled 
regime is presented.
This method involves a new type of weak coupling expansion
which is multiplicative rather than additive in nature
and allows perturbative calculation of
partition function zeroes.
Application of the method to the single flavour
Gross-Neveu model gives a phase diagram
consistent with the parity symmetry breaking scenario
of Aoki and provides new quantitative information on the 
width of the 
Aoki phase in the weakly coupled sector.

\end{abstract}

\end{titlepage}

%%%%%%%%%%%%%%%%%%%%%%%%%%%%%%%%%%%%%%%%%%%%%%%%%%%%%%%%%%%%%%%%%%%%%%%%
%%%%%%%%%%%%%%%%%%%%%%%%%%%%%%%%%%%%%%%%%%%%%%%%%%%%%%%%%%%%%%%%%%%%%%%%
%%%%%%%%%%%%%%%%%%%%%%%%%%%%%%%%%%%%%%%%%%%%%%%%%%%%%%%%%%%%%%%%%%%%%%%%
%%%%%%%%%%%%%%%%%%%%%%%%%%%%%%%%%%%%%%%%%%%%%%%%%%%%%%%%%%%%%%%%%%%%%%%%

\setcounter{equation}{0}

In lattice field theory, there has been considerable discussion
on the phase diagrams of theories with Wilson fermions
(see, e.g., 
\cite{Kawamoto,Aoki,support1,support2,support3,Sharpe,us,
against,more,Creutz,width}).
These can be considered as
 statistical mechanical systems, and have
 rich phase structures
whose existence is due to lattice artefacts.
The Wilson fermion 
hopping parameter is $\kappa = 1/2({\hat{M}_0}+d)$
 where ${\hat{M}_0}$ is the dimensionless bare fermion mass
and $d$ the lattice dimensionality.
It is well known that a 
system of free Wilson fermions exhibits a  phase 
transition
at $\kappa= 1/(2d)$
and that massless fermions appear at this point in the continuum limit.
Discussions concern the extent to which this phase
transition persists in the presence of a bosonic field.
In QCD, where Wilson fermions couple to gauge fields with a strength
given by the dimensionless coupling ${\hat{g}}$,
there are two candidates for the phase diagram.
In the first, pioneered by Kawamoto,  
the expectation is that 
there is a line of phase transitions (the ``chiral line'') 
extending from the strong coupling limit  to the weakly coupled 
one and along which the pion and quark  masses
vanish \cite{Kawamoto}. 
Such vanishing is symptomatic 
 of spontaneous chiral symmetry breaking.
Approaching the continuum limit, at ${\hat{g}}=0$,
along the chiral line in particular
is then expected to recover massless physics.
This is still sometimes referred to as the
`conventional' picture.

The second candidate
 phase diagram for QCD  was determined by Aoki on the basis of
comparison with the Gross-Neveu model \cite{Aoki}. 
The Gross-Neveu  model serves as a prototype for QCD \cite{GN}. Indeed,
except for confinement,  it
has features similar to QCD. One of
these features is asymptotic freedom, so that in the Gross-Neveu model,
as in QCD, the continuum limit is taken in the weakly coupled zone.
Two features distinguish 
Aoki's phase diagram from the earlier `conventional' picture.
Firstly, instead of a single critical line,
Aoki's analysis advocates 
the existence of two lines 
extending from the
strongly to weakly coupled limits, 
with a number of critical points
at ${\hat{g}} = 0$
linked by cusps (see Fig.~1).
The region above the cusps and between the two extended lines
is often referred to as the Aoki phase.
Secondly, the existence of the phase transition in Aoki's scenario is 
due
to spontaneous 
parity symmetry breaking within the Aoki phase,
as opposed to chiral symmetry breaking.
(Indeed, Wilson fermions explicitly violate chiral symmetry.)
This is signaled by a non zero vacuum expectation value
of the pseudoscalar
operator $\pi = {\bar{\psi}} i \gamma_S \psi$ in the thermodynamic limit.
The masslessness of the  pion is then attributed to the divergence
of a correlation length associated with this second order phase
transition. 
In the multiflavour
 case, flavour symmetry is also broken in the Aoki phase
since the pion, whose expectation value is nonvanishing, also carries
flavour.
The continuum limit has to be approached from outside the Aoki
phase since parity and flavour are conserved in the strong interaction.
The physical meaning of an approach to the continuum limit
from within the Aoki phase is unclear \cite{Aoki}.
There exists substantial evidence supporting Aoki's scenario in the 
strongly
coupled regime \cite{Aoki,support1,support2,support3,Sharpe,us}. 
In the weakly coupled regime the evidence has, however, been 
controversial \cite{against} (see \cite{Sharpe} for recent discussions
on this topic).

In asymptotically free theories the weakly coupled
region is the appropriate one for  the continuum limit.
Recently, Creutz \cite{Creutz} questioned whether
the Aoki phase, pinched between the arms of  cusps,
is ``squeezed out'' at non-zero coupling or whether
 it only vanishes in the weak coupling limit 
(see, also, \cite{Sharpe,width}).

The purpose of this paper is twofold. We introduce a new type of
expansion which is multiplicative rather than additive in nature
and from which information on the partition function zeroes
of the theory can be extracted in a rather natural way \cite{us}. 
Secondly, we address the question of the ``squeezing out'' of
the Aoki phase at weak coupling.
This multiplicative approach to the single flavour Gross-Neveu model,
shows that the width of the central  Aoki cusp
is   ${\cal{O}}(\hat{g}^2)$
while the Aoki phase has not yet emerged at this order
from the left and right extremes.

The Gross-Neveu model is actually a two dimensional
model of fermions only, which interact
through a short range quartic interaction \cite{GN}.
In Euclidean continuum space, the model  with
a single fermion flavour is given by the
four--fermi action
\begin{equation}
 S_{ \rm{GN} }^{ \rm{(cnm)} }
=
\int{
     d^2x 
     \left\{
                     \bar{\psi}(x) 
                     \left( 
                           \partial \!\!\!/ 
                           \:+ M 
                     \right) 
                     \psi (x)
                   -
                   \frac{g^2}{2}
                   \left[
                         \left(
                                      \bar{\psi}(x) \psi (x)
                         \right)^2
                         +
                         \left(
                                      \bar{\psi} (x)i \gamma_S \psi (x)
                         \right)^2
                   \right]
      \right\}
}
\quad ,
\label{cfIz2.1cfHa1.1}
\end{equation}
where
$
\gamma_S 
 =  
 i^{-1}
 \gamma_1  \gamma_2
$
and the fermion field has 2 spinor components.
Bosonizing the action gives for the partition function,
$
 Z_{GN}^{({\rm{cnm}})} =
\int{
 {\cal{D}} \phi
 {\cal{D}} \pi 
 {\cal{D}} \bar{\psi}
 {\cal{D}} \psi
 \exp{(-S)}
}
$,
where
\begin{equation}
 S
=
\int{
     d^2x \left\{
                 \bar{\psi} \left( \partial \!\!\!/ 
                         \:+ M \right) \psi
                +
                \frac{1}{2g^2}
                \left[
                      \phi^2
                      + \pi^2
                \right]
               +
                     \phi
               \bar{\psi} \psi
               +
               \pi
               \bar{\psi} i \gamma_S \psi
       \right\}
}
\quad ,
\end{equation}
and where $\phi (x)$ and $\pi (x)$ are auxiliary boson fields.
The corresponding Wilson action in terms of dimensionless lattice 
quantities is 
$ S_F^{  ({\rm{W}})  }  =  S_F^{(0)} + S_{  ({\rm{int}})  } 
+ S_{  ({\rm{bosons}})  }$,
where \cite{support1}
\begin{equation}
 S_F^{(0)}
 =
 \frac{1}{2\kappa}
 \sum_n{\bar{\psi}(n)\psi(n)}
 -
 \frac{1}{2}\sum_{n,\mu}{
 \big\{
         \bar{\psi}(n)
         ( 1 - \gamma_\mu )
         \psi(n+\hat{\mu})
 }
    + 
          \bar{\psi}(n+\hat{\mu})
         ( 1 + \gamma_\mu )
         \psi(n)
 \big\}
 ,
\label{cfRo14.2d}
\end{equation}
\begin{equation}
 S_{  ({\rm{int}})  } 
=
{\hat{g}} \sum_n{\phi(n) \bar{\psi}(n) \psi(n)
}	
+
{\hat{g}} \sum_n{\pi(n) \bar{\psi}(n) i \gamma_S \psi(n)
}	
\quad,
\end{equation}
and
\begin{equation}
 S_{  ({\rm{bosons}})  } 
=
 \frac{1}{2}
 \sum_n{ \left[\phi^2(n) + \pi^2(n) \right]
}	
\quad,
\end{equation}
and where the auxiliary fields have been rescaled $\phi \rightarrow 
{\hat{g}} \phi$, $\pi \rightarrow {\hat{g}} \pi$ to explicitly 
display
the order of the interactive part of the action.
Here, lattice sites are labeled 
$ n_\mu = -N/2, \dots, N/2 - 1$, where $N$ is the 
 number of sites in each of the two directions. 
We assume $N$ is even.

Using the lattice Fourier transform,
$\psi(n) = (1/Na)^2\sum_k{\psi(k)\exp{(ikna)}}$,
where $a$ is the lattice spacing,
the fermionic action can be written
\begin{equation}
 S_F^{(0)}+ S_{  ({\rm{int}})  } 
 = 
 \frac{1}{N^2 
 a^{4}}\sum_{q,p}{\bar{\psi}(q)M^{({\rm{W}})}(q,p)\psi(p)}
\quad,
\label{pageGN3}
\end{equation}
where the $2 N^2 \times 2 N^2$ fermion matrix is
$
M^{({\rm{W}})}(p,q)=M^{(0)}(p,q)+M^{({\rm{int}})}(p,q)
$,
with free part
\begin{equation}
M^{(0)}(q,p)
 =
 \delta_{qp}
 \left[
  \frac{1}{2\kappa} - \sum_\mu(\cos{p_\mu a}
                       - i \gamma_\mu \sin{p_\mu a} )
 \right]
\quad ,
\end{equation}
and interactive part
\begin{equation}
M^{({\rm{int}})}(q,p)
=
\frac{{\hat{g}}}{N^2}
\sum_n{
e^{i (p-q) n a}
 \left[
  \phi(n) + \pi (n) i \gamma_S
 \right]
}
\quad .
\end{equation}

It is appropriate to impose antiperiodic boundary 
conditions in the temporal ($1$-) direction and periodic boundary 
conditions in the spatial ($2$-) direction in coordinate space. 
With these mixed boundary conditions the momenta for the
Fourier transformed fermion fields  are
$ p_\mu = 2\pi {\hat{p}}_\mu /Na $,
where $ {\hat{p}}_1 \in \{ -N/2+1/2,  \dots, N/2 - 1/2\}$ and
$ {\hat{p}}_2  \in \{  -N/2,  \dots, N/2 - 1$\}.
Integration over the Grassmann variables gives the full
partition function 
\begin{equation}
Z = 
\int{
 {\cal{D}} \phi
 {\cal{D}} \pi 
 {\cal{D}} \bar{\psi}
 {\cal{D}} \psi
 \exp{\left(-S_F^{\rm{(W)}}\right)}
}
\propto \left\langle 
 \det{  M^{  ({\rm{W}})  }  } \right\rangle
\propto \left\langle 
  \prod_{\alpha,p} \lambda_\alpha(p)  \right\rangle
\quad ,
\label{zzzz}
\end{equation}
with $\lambda_\alpha(p)$ the eigenvalues of the fermion matrix
and the expectation
values being taken over the bosonic fields.

In the free field case the eigenvalues of $M^{(0)}$ are easily 
calculated
and found to be
\begin{equation}
 \lambda^{(0)}_\alpha(p)
 =
 \frac{1}{2 \kappa}
 -
 \eta^{(0)}_\alpha(p)
\quad,
\label{eigenvalues}
\end{equation}
where 
\begin{equation}
\eta^{(0)}_\alpha(p)
=
 \sum_{\mu=1}^2  \cos{p_\mu a} 
 - i (-)^\alpha  \sqrt{ \sum_{\mu=1}^2{\sin^2{p_\mu a}}}
\label{free}
\end{equation}
are the Lee--Yang zeroes of the free theory \cite{LY}.
Note that the eigenvalues (\ref{eigenvalues}) and the zeroes
(\ref{free}) are degenerate with respect to 
$p_\mu \rightarrow - p_\mu$.
Furthermore,
 the lowest zeroes in the free case, and those responsible for the
onset of critical behaviour, are two-fold degenerate in two dimensions.
These lowest zeroes are 
$\eta_\alpha (\pm |p_1|,p_2)$ where $|{\hat{p}}_1| = (N-1)/2$ or $1/2$ 
and ${\hat{p}}_2 = -N/2$ or $0$ and 
impact on the real $1/2\kappa$ axis at $-2$, $0$ and $2$.
Finally note that the zeroes in the upper half plane
are given by $\alpha = 1$, while their complex conjugates correspond
to $\alpha = 2$.

The standard additive weak coupling expansion of the full fermion
determinant is the Taylor expansion of
\begin{eqnarray}
 \det{ M^{ ( { \rm{W} } ) } }
 &=&
 \det{ M^{(0)} }
 \times 
 \det \left({ M^{(0)} }^{-1}
 M^{ ( {\rm{W}} ) }\right)
\nonumber \\
~ & = &
 \det{ M^{(0)} }
  \exp{ 
       {\rm{tr}}
       \ln{
            \left(
                  1+{M^{(0)}}^{-1}
                    M^{ ( { \rm{int} } ) }
            \right)
          }
      }
\quad .
\end{eqnarray}
This expansion is
\begin{equation}
 \frac{  \det{M^{( {\rm{W}} )}}
      }{
         \det{M^{(    0     )}}
      }
  =
  1 
  + 
  \sum_{i=1}^{2N^2}{
         \frac{ M^{({\rm{int}})}_{ii}
              }{
                \lambda_i^{(0)}
              }
        }
  -\frac{1}{2}
   \sum_{i,j=1}^{2N^2}{
              \frac{ M^{({\rm{int}})}_{ij} 
                     M^{({\rm{int}})}_{ji}
                   }{
                     \lambda_i^{(0)}
                     \lambda_j^{(0)}
                    }
             }
 +\frac{1}{2}
  \sum_{i,j=1}^{2N^2}{
              \frac{ M^{({\rm{int}})}_{ii} 
                     M^{({\rm{int}})}_{jj}
                   }{
                     \lambda_i^{(0)}
                     \lambda_j^{(0)}
                    }
             }
+ \dots
\quad ,
\label{dia}
\end{equation}
where the indices $i$ and $j$ stand for the combination of Dirac
index and momenta $ ( \alpha, p ) $ which 
label fermionic matrix elements, so that $M_{ij}^{({\rm{int}})}$
represents $\langle \lambda_\alpha^{(0)} (p) | M^{({\rm{int}})}(p,q)
| \lambda_\beta^{(0)} (q) \rangle $.
Here   $| \lambda_\beta^{(0)} (q) \rangle $ 
represents a free fermion 
eigenvalue.
The traces in (\ref{dia}) are just the diagrams which contribute to
the vacuum polarization tensor.

Setting $ t_i =  \langle M_{ii}^{({\rm{int}})} \rangle $
and 
$ t_{ij} = t_{ji} 
= \langle M_{ij}^{({\rm{int}})} M_{ji}^{({\rm{int}})}\rangle 
-
\langle M_{ii}^{({\rm{int}})} M_{jj}^{({\rm{int}})}\rangle 
$, 
the ratio of partition functions is, from (\ref{dia}),
\begin{equation}
\left\langle 
\frac{ 
       \det{M^{({\rm{W}})}   } 
      }{
       \det{M^{(0)}}
      }
\right\rangle 
 = 
 1
 +
 \sum_{i=1}^{2N^2}{
                   \frac{t_i}{\lambda_i^{(0)}}
                  } 
-\frac{1}{2}  
 \sum_{i,j=1}^{2N^2}{
                      \frac{
                            t_{ij}
                          }{
                            \lambda_i^{(0)}
                            \lambda_j^{(0)}
                           } 
                     }
 + \dots
 .
\label{additive}
\end{equation}
We note at this point that this expansion is analytic in $1/2\kappa$ 
with
poles at $1/2\kappa = \eta_i^{(0)}$. 

For the Gross-Neveu model, 
calculation of the
pure bosonic expectation values is particularly
simple.
Indeed, one has that $\langle \phi (n) \rangle = 
\langle \pi (n) \rangle = 0$ and
\begin{equation}
 \langle \phi (n) \phi(m)\rangle = \langle \pi(n) \pi (m) \rangle 
=
2 
\delta_{nm}
\quad .
\end{equation}
The required bosonic  expectation values of the matrix elements are
found to be
\begin{eqnarray}
 t_i & \equiv & t_{\alpha, p} = 0
\quad ,
\label{a} \\
 t_{ij} & \equiv  & t_{(\alpha, p) (\beta,q) } 
 = \frac{2\hat{g}^2}{N^2} 
\left\{
  (-1)^{\alpha+\beta}
 \frac{\sum_\rho 
                \sin{q_\rho}
                \sin{p_\rho}
     }{
       \sqrt{
             \sum_\mu
                      \sin^2{q_\mu}
             \sum_\nu
                      \sin^2{p_\nu}
            }
       }
 -1
\right\}
\quad .
\label{c} 
\end{eqnarray}

An alternative formulation of the partition function may be obtained
by writing the Wilson fermion matrix as $M^{({\rm{W}})} = 1/2\kappa
+ H $ where $H$ is the hopping matrix. The fermion
determinant $\det M^{({\rm{W}})} = \det(1/2\kappa + H)$, is a polynomial
in $1/2\kappa$ since for finite lattice size these matrices are
of finite dimension. Indeed, for an $N\times N$ lattice
this polynomial is of degree $2N^2$.
Therefore the  bosonic expectation value
of the fermion determinant is also a polynomial 
of the same degree in $1/2\kappa$ and as such
may be written in terms of its $2N^2$ zeroes, now labeled $\eta_i$.

We may thus write a `multiplicative' weak coupling expansion as
\begin{equation}
     \left\langle 
 \frac{ 
      \det{M^{({\rm{W}})}}
      }{
       \det{M^{(0)}}
      }
 \right\rangle 
= 
 \prod_{i=1}^{2N^2}
\left( \frac{1/2\kappa - \eta_i}{\lambda_i^{(0)}} \right)
 = 
\prod_{i=1}^{2N^2}\left(
  1 - \frac{\Delta_i}{\lambda_i^{(0)}}
                    \right)
 ,
\label{multiplicative}
\end{equation}
where $\Delta_i = {\eta}_i - {\eta_i}^{(0)}$ are the shifts that 
occur
in the zeroes when the bosonic fields are turned on.
These are the quantities to be determined.
Note that the expression (\ref{multiplicative}) is,
like (\ref{additive}),
analytic in $1/2\kappa$ with
poles at $\eta_i^{(0)}$. Expanding (\ref{multiplicative}) gives
\begin{equation}
     \left\langle 
 \frac{ 
      \det{M^{({\rm{W}})}}
      }{
       \det{M^{(0)}}
      }
 \right\rangle 
= 1 -
 \sum_{i=1}^{2N^2}{
 \frac{\Delta_i}{\lambda_i^{(0)}}
}
+
\frac{1}{2} \sum_{i=1}^{2N^2}{\sum_{j\ne i}^{2N^2}{
 \frac{\Delta_i \Delta_j}{\lambda_i^{(0)}\lambda_j^{(0)}}
}}
+ \dots
\quad.
\label{expmult}
\end{equation}

In the free fermion theory, the eigenvalues and zeroes 
of (\ref{eigenvalues}) and (\ref{free}) are
two- or four- fold  degenerate with respect to momentum inversion. 
Let $\{n\}$ denote the $n^{\rm{th}}$ 
degeneracy class, so that the $D_n$ eigenvalues
$\lambda^{(0)}_{n_1} = \dots
 =\lambda^{(0)}_{n_{D_n}}$ are identical to $\lambda^{(0)}_{n}$, say.
Let $1/2\kappa = \eta_n^{(0)} + \epsilon$ and expand the
additive and multiplicative expressions (\ref{additive}) and
(\ref{expmult}) order by order in $\epsilon^{-1}$.
Indentification of the expansions
 yields relationships between the known quantities $t_i$, 
and $t_{ij}$ and the shifts in the positions of the
zeroes, $\Delta_i$, to
 ${\cal{O}}(\epsilon^{-1})$ and ${\cal{O}}(\epsilon^{-2})$.
The ${\cal{O}}(\epsilon^{-1})$ relationship is
\begin{equation}
 \sum_{ 
       n_i \in \{ n \}
       }{
          \Delta_{n_i}
          \left\{
                 1 - 
                 \sum_{ j \not\in \{n\} }{
                                         \frac{\Delta_j
                                              }{
                                                \eta_n^{(0)} 
                                              - \eta_j^{(0)}
                                              }
                                        }
          \right\}
        }
 = 
 \sum_{ 
       n_i \in \{ n \}
      }{
        \sum_{
               j\not\in \{n\}
             }{
                \frac{
                      t_{n_i j}
                      }{
                      \eta_n^{(0)} - \eta_j^{(0)}
                      }
             }
       }
\quad ,
\label{ordere1}
\end{equation}
having used (\ref{a}), 
while that to ${\cal{O}}(\epsilon^{-2})$ is
\begin{equation}
\sum_{n_i,n_j \in \{n\},n_i \ne n_j}{
    \Delta_{n_i} \Delta_{n_j} }
= - \sum_{n_i,n_j \in \{n\}}{
t_{n_i n_j}}
\quad .
\label{ordere2}
\end{equation}

These  relationships can be considered order by order in the
coupling as well. Let
$\Delta_i = {\eta}_i^{(1)} + {\eta}_i^{(2)}
+ {\cal{O}}({\hat{g}}^3) $, where 
${\eta}_i^{(1)}$ and
and ${\eta}_i^{(2)}$
are, respectively,  the order ${\hat{g}}$ 
and order ${\hat{g}}^2$ shifts in the $i^{\rm{th}}$ zero.
One finds that the ${\cal{O}}(\epsilon^{-1})$ equation to order $\hat{g}$
is
\begin{equation}
 \sum_{n_i \in \{n\}}{ {\eta}_{n_i}^{(1)} }  = 0
\quad ,
\label{M27.1}
\end{equation}
while to  order $\hat{g}^2$ it is
\begin{equation}
 \sum_{n_i \in \{n\}}{ {\eta}_{n_i}^{(2)} }  =  
  \sum_{n_i \in \{n\},j \not\in \{n\} }{
 \frac{t_{n_ij}
  }{\eta_n^{(0)}-\eta_j^{(0)}}
}
\label{M27.2}
\quad .
\end{equation}
Also, the ${\cal{O}}(\epsilon^{-2})$ equation, which
is entirely ${\cal{O}}(\hat{g}^{2})$, is
\begin{equation}
 \sum_{
        n_i \in \{n\}
      }{
             \left(
               {\eta_{n_i}}^{(1)}
         \right)^2
      } 
  = 
 \sum_{
       n_i,n_j \in \{n\}
      }{
                 t_{n_i n_j}
       }
\label{M27.3}
\quad .
\end{equation}
With relations (\ref{M27.1})-(\ref{M27.3}), 
the multiplicative expression (\ref{expmult}) recovers (\ref{additive})
to ${\cal{O}}(\hat{g}^{2})$.
Now the additive and multiplicative expressions (\ref{additive}) and
(\ref{multiplicative})  coincide to
${\cal{O}}(\hat{g}^{2})$  everywhere in the complex hopping
parameter plane and  arbitrarily close to any pole. 

In the free case, the zeroes responsible for criticality are two
fold degenerate. For weak enough coupling, 
one expects these zeroes to govern critical behaviour
in the presence of weakly
coupled bosonic fields too.
From (\ref{M27.1}) and (\ref{M27.3}), the 
first order shifts to two-fold degenerate zeroes are
\begin{equation}
 \eta^{(1)}_{n_i} =
 \pm
 \sqrt{
       t_{n_1 n_2}
      }
\quad ,
\label{1st}
\end{equation}
where $n_i \in \{ n \}$ for $i =1$ or $2$.
The second order equation in the two-fold degenerate case is
\begin{equation}
\eta^{(2)}_{n_1}
+  \eta^{(2)}_{n_2}
 = 
\sum_{j\not\in \{n\}}{
                \frac{
                             t^{(2)}_{jn_1}
                           + t^{(2)}_{jn_2}
                     }
                     {
     \eta_n^{(0)} - \eta_j^{(0)}
                     }
                      }
\quad .
\end{equation}
Removing the bosonic field expectation values converts the
problem into the determination of the eigenvalues of a weakly
perturbed matrix whose free eigenvalues are two-fold degenerate.
More explicitly, with boson expectation values removed, the 
eigenvalues
are $\lambda_i = \lambda_i^{(0)} - \eta_i^{(1)}-\eta_i^{(2)}$,
which may be determined from conventional  time independent 
perturbation theory. This condition yields enough to
fully determine the zeroes to order $\hat{g}^2$. 
Indeed, the second order shifts are
\begin{equation}
  \eta^{(2)}_{n_i}
= 
\sum_{j\not\in \{n\}}{
                      \frac{
                             t_{jn_i}
                           }{
               \eta_n^{(0)} - \eta_j^{(0)}
                            }
                      }
\quad .
\label{2nd}
\end{equation}

Now using  (\ref{c}),
the ${\cal{O}}(\hat{g})$ and ${\cal{O}}(\hat{g}^2)$   
shifts
 for the erstwhile two-fold degenerate zeroes,
$\eta_\alpha(\pm|p_1|,p_2)$ (for $\hat{p}_2= 0$ or
 $-N/2$), are, respectively,
\begin{eqnarray}
 \eta^{(1)}_\alpha(\pm|p_1|,p_2)
 & = &
 \pm
 i \frac{2g}{N}
\quad ,
\label{1c}
\\
 \eta^{(2)}_\alpha(\pm|p_1|,p_2)
 & = &
- \frac{2g^2}{N^2}
\sum_{(\beta,q)\not\in\{(\alpha,p)\}}{\frac{1}{\eta^{(0)}_\alpha(p)
-\eta^{(0)}_\beta(q)
}}
\quad .
\label{2c}
\end{eqnarray}

The partition function zeroes are `protocritical points'
\cite{proto} whose real parts are pseudocritical 
points. In the thermodynamic limit these  become the true
critical points of the theory and their determination amounts to 
determination of the weakly coupled 
phase diagram, the critical line being traced out
by the impact of zeroes on to the real hopping parameter axis. 
Thus, the phase diagram is given to order ${\hat{g}}^2$ by the 
limit
\begin{equation}
\frac{1}{2\kappa} = \lim_{N\rightarrow \infty}{\left\{
\eta_\alpha^{(0)}(p)
+
\eta_\alpha^{(1)}(p)
+
\eta_\alpha^{(2)}(p)
\right\}}
\quad ,
\end{equation}
where $p$ is the momentum corresponding to the lowest zeroes.

At order ${\hat{g}}^0$, the zeroes (\ref{free})
impact on the real $1/2\kappa$ axis at $-2$, $0$ and $2$,
giving three different continuum limits, corresponding 
to the nadirs of
the three Aoki cusps \cite{Aoki}. The true continuum
limit is $1/2\kappa = 2$.
From (\ref{1c}) and (\ref{2c}), 
the ${\cal{O}}({\hat{g}}^2)$-shift 
is the shift in the average position
of the two zeroes while their relative separation is 
${\cal{O}}({\hat{g}})$.
In the thermodynamic limit, the ${\cal{O}}({\hat{g}})$ terms 
in (\ref{1c}) vanish.
One finds, numerically, that the imaginary contribution to
the ${\cal{O}}({\hat{g}}^2)$ term (\ref{2c})
also vanishes while
the real part becomes an $N$-independent constant.
Indeed, the factor
\begin{equation}
\frac{1}{N^2}
\sum_{(\beta,q)\not\in\{(\alpha,p)\}}{\frac{1}{\eta^{(0)}_\alpha(p)
-\eta^{(0)}_\beta(q)
}}
\label{factor}
\end{equation}
approaches approximately $0.77$ and $-0.77$ for
$(\hat{p}_1,\hat{p}_2) = (\pm1/2,0)$
and 
$(\pm(N-1)/2,-N/2)$ respectively,
and corresponding to the rightmost and leftmost critical lines.
Also, (\ref{factor}) is
approximately $0.2$ and $-0.2$ for
$(\hat{p}_1,\hat{p}_2) = (\pm(N-1)/2,0)$
and 
$(\pm 1/2,-N/2)$ respectively, these two lines generating
the inner cusp.

Therefore, the degeneracy of the free fermion critical point 
corresponding to the central  cusp in Aoki's phase diagram
is indeed lifted and two critical lines emerge in the
presence of weak bosonic coupling.
These critical lines are $1/2 \kappa \simeq \pm 0.4 \hat{g}^2$.
The Aoki phase does not yet emerge to ${\cal{O}}({\hat{g}}^2)$
from the left- and rightmost critical points.
This is the answer to the question posed by Creutz in \cite{Creutz}
at least for the single flavour Gross-Neveu model.
 Instead the 
right- and leftmost critical lines are 
$1/2 \kappa \simeq \pm( 2- 1.54 \hat{g}^2)$ respectively.

\begin{figure}[htb]
\vspace{6.5cm}               
\includegraphics{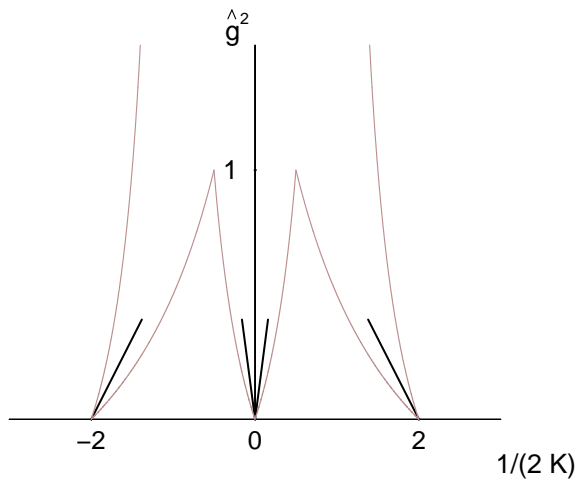}
\caption{The phase diagram for the
Gross-Neveu model 
in the weakly coupled region (to ${\cal{O}}(\hat{g}^2)$)
(dark lines) and  a schematic representation of
the expected Aoki phase diagram (light curves).}
\label{fig}
\end{figure}

This weakly coupled
phase diagram is pictured in Fig.~1 (dark lines). 
From left to right, the critical lines are traced out by
the thermodynamic limits of the zeroes indexed by
$(\hat{p}_1,\hat{p}_2)$ =
$(\pm(N-1)/2,-N/2)$,
$(\pm(N-1)/2,0)$,
$(\pm 1/2,-N/2)$ and
$(\pm 1/2,0)$
respectively. The lighter curves are 
a schematic representation of
the expected full phase diagram.

In conclusion, we have developed a new type of weak coupling
expansion which is multiplicative rather than additive in
nature and focuses on the Lee-Yang zeroes, or protocritical points,
of a lattice field theory with Wilson fermions. 
This expansion is applied to 
the Gross-Neveu model, where the existence of an Aoki
phase was first suggested. 
The weakly coupled regime is the one
of primary interest as it is there, as with 
all asymptotically free
models, that the continuum limit is taken.
The method, applied to the single flavour Gross-Neveu model,
yields a phase diagram in this 
region which is consistent with
that of Aoki and the widths of 
the Aoki cusps are determined to order ${\hat{g}}^2$.

{\bf{Acknowledgements:}} RK wishes to thank M. Creutz for a 
discussion.
 
%---------------------------------------------------------

\end{document}